\documentclass[journal]{IEEEtran}
\usepackage{cite}
\usepackage{amsmath,amssymb,amsfonts}
\usepackage{hyperref}
\usepackage{epstopdf}
\usepackage{algorithm}
\usepackage{algorithmicx}
\usepackage{graphicx}
\usepackage{graphics}
\usepackage{textcomp}
\usepackage{xcolor}
\usepackage{pdfpages}
\usepackage{float}
\usepackage{placeins}
\usepackage{optidef}
\usepackage{ dsfont }
\usepackage{setspace}
 \usepackage{float}
 \usepackage{epsfig,color,algorithm}
 \usepackage{algpseudocode}
 \usepackage{multirow}
\usepackage{changes}

\ifCLASSINFOpdf
  
\else
 
\fi
  
\usepackage{epsfig} 
\ifCLASSOPTIONcompsoc
  \usepackage[caption=false,font=normalsize,labelfont=sf,textfont=sf]{subfig}
\else
  \usepackage[caption=false,font=footnotesize]{subfig}
\fi

\usepackage{fixltx2e}
\usepackage{stfloats}

\usepackage{soul}
\usepackage{color}

\begin{document}

\title{Identification of The Number of Wireless Channel Taps Using Deep Neural Networks}


\author{Ahmad M. Jaradat, Khaled Walid Elgammal, Mehmet Kemal \"{O}zdemir
        and H\"{u}seyin Arslan,~\IEEEmembership{Fellow,~IEEE}
\thanks{Ahmad M. Jaradat and Khaled Walid Elgammal are with the Department of Electrical and Electronics Engineering, Istanbul Medipol University, Istanbul 34810, Turkey (email: [ahmad.jaradat,khaled.abdelfatah]@std.medipol.edu.tr). 

Mehmet Kemal \"{O}zdemir is with the Department of Electrical and Electronics Engineering, Istanbul Medipol University, Istanbul 34810, Turkey (email: mkozdemir@medipol.edu.tr). 

H\"{u}seyin Arslan is is with the Department of Electrical and Electronics Engineering, Istanbul Medipol University, 34810 Istanbul, Turkey, and also with the Department of Electrical Engineering, University of South Florida, Tampa, FL 33620 USA (email: huseyinarslan@medipol.edu.tr).}}

\maketitle


\begin{abstract}
In wireless communication systems, identifying the number of channel taps offers an enhanced estimation of the channel impulse response (CIR).
In this work, {efficient} identification of the number of {wireless} channel taps has been achieved {via deep neural networks (DNNs)}, 
{where we} modified an existing DNN and analyzed its convergence performance using only the transmitted and received signals of a wireless system.
The displayed results demonstrate that the {adopted} DNN {accomplishes superior} performance in identifying the number of channel taps, {as compared to an existing algorithm called Spectrum Weighted Identification of Signal Sources (SWISS)}.
\end{abstract}


\IEEEpeerreviewmaketitle


\section{Introduction}
\IEEEPARstart{T}{he} {identification of the number of channel taps is a challenging task in real-time with fast channel variations.}
{Precise channel estimation with a {better} restoration of transmitted signals is possible by identifying the number of taps for unknown environments with no prior information.}
{Therefore, an effective technique to recognize these taps is needed.}
Various wireless channel identification techniques, {like} \cite{Xiao2014TWC,Dong2015CC,Dengao2017BP}, {assume prior} information at the receiver about the multipath delay profile for the channel. However, this profile is usually unknown in practical use cases.

Motivated by {the potential improvements that can be accomplished over the state-of-the-art}, {this paper proposes a} deep neural network (DNN) to identify a sparse channel parameter called the number of taps from transmitted and received data of a wireless system.
Our proposed identification technique enhances {the} spectral efficiency of the communication system since {the adopted} DNN requires no extra {transmitted signals} {other than} the {channel input and output data}. 

Deep learning (DL) can generate extremely high-level data representations from a massive amount of data. 
More specifically, the DNNs can create maximally sparse representations {of the signals of interest}.
{To reflect our approach of DNN modeling, we used} the basic network {in \cite{Xin2016} for finding the maximum sparsity.}
{The DNN in \cite{Xin2016} unfolds the high-complexity Iterative Hard Thresholding (IHT) algorithm given in \cite{IHT2009}. 
{The research study \cite{Xin2016} shows the superior theoretical and empirical performance of this DNN compared to the IHT method.}
Also, the DNN promotes the maximal sparsity of its input with incomparable complexity.} 


{To highlight the advantages of the adopted DNN in the proposed identification solution, we chose to compare its performance with an existing algorithm called Spectrum Weighted Identification of Signal Sources (SWISS) \cite{8377366,9107261}.} {The SWISS algorithm solves the identification problem of path-number by determining the optimum combination of the discrete Fourier transform (DFT) components in the weighted DFT of the received signal. The resultant signal represents the reconstructed channel signal, which has the minimum Euclidean distance to the received signal. The number of channel paths is then identified by calculating the number of significant components in the weight vector. Similar to the proposed DNN-based method, the SWISS algorithm does not need prior information about the number of channel taps.} 

{Unlike the existing SWISS algorithm, our main contributions to the literature are summarized as follows:}
\begin{itemize}
    \item {A new DNN-based technique is proposed to solve the {multi-label classification problem} of identifying the number of wireless channel taps.} In {the proposed technique, prior information is assumed to be not available.} 
    \item A measurement-based simulator has been used for channel data generation, which {is used} for training, validating, and testing the {adopted} DNN. 
    \item The proposed DNN-based technique has significantly improved accuracy performance over the SWISS algorithm \cite{8377366} while maintaining moderate complexity.
\end{itemize}

The rest of this {paper} is organized as follows. Section II provides a summary of the fundamental system model. Section III presents the datasets generation and simulation setup. {The simulation results are presented in Section IV}. Finally, Section V provides a conclusion {with} some future directions. 

\section{System Model}
The deterministic parameters of wireless channels include the number of taps, taps delays, and delay spread. 
Different taps of the channel impulse response (CIR) experience {pathloss} and time-variations in a multipath radio environment. The CIR could be formulated as \cite{Idrees2015}

\begin{equation} 
h(\tau, t)=\sum_{l=0}^{L-1} \beta_{l}\thinspace \mu_{l}(t)\thinspace \delta\left(\tau-\tau_{l}\right),
\label{eq1}
\end{equation}
where $L$ represents the total number of channel taps, $\beta_{l}$ and $\tau_{l}$ represent {pathloss} and delay of $l$-th channel tap, respectively. $\mu_{l}(t)$ represents the time variation of the $l$-th channel tap{, and $\delta(.)$ is the delta function}. 

{Generally, double-directional CIR is adopted as a fundamental deterministic description of the channel in modern wireless communication systems. This CIR can be formulated as \cite{spatial2014}}

\begin{equation} 
h(\tau, t, \psi, \theta)=\sum_{l=0}^{L-1} \beta_{l} \mu_{l}(t) \delta\left(\tau-\tau_{l}\right) \delta\left(\theta-\theta_{l}\right) \delta\left(\psi-\psi_{l}\right),
\label{mmWaveeq}
\end{equation}
{where $\theta$ and $\psi$ represent the angle of departure and angle of arrival, respectively.}

\begin{figure*}
\centering

\includegraphics[width=6in]{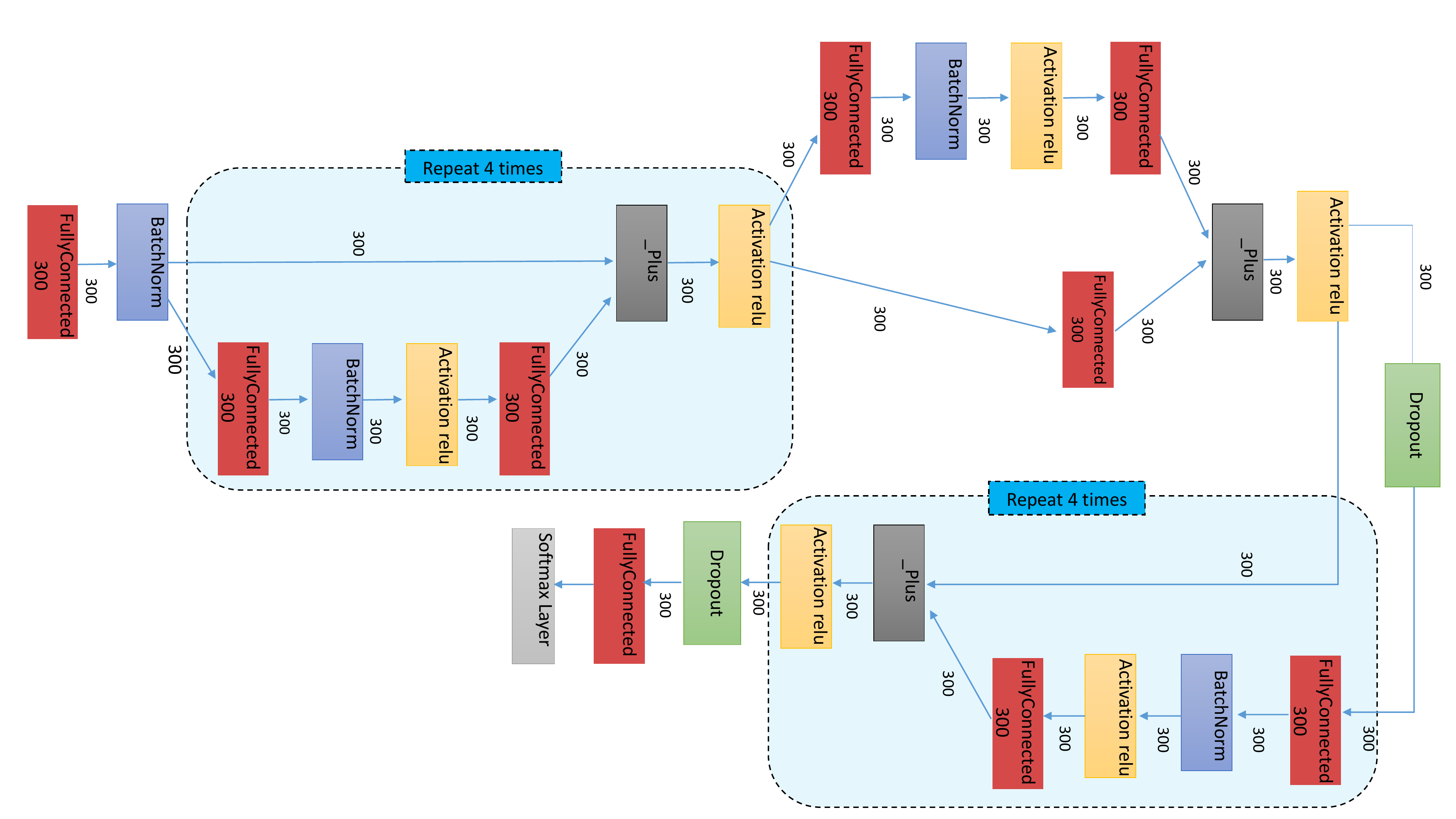}

\caption{Basic network structure \cite{Xin2016}.}
\label{networkfig}
\end{figure*} 

The received signal {faded by a multipath channel}, {assuming a noise-free system for simplification, can be represented by}

\begin{equation} 
y(t)=x(t) \ast h(\tau, t, \psi, \theta),
\label{eq2}
\end{equation}
where $\ast$ represents the convolution operator {and $x(t)$ is the transmitted signal.}

With different {wireless} environments, {the parameter $L$} varies due to different propagation paths. {Therefore,} it is necessary to identify $L$ in the CIR {since it is usually unknown in practical scenarios}. 

{The input samples to the DNN are represented as}
\begin{equation} 
\mathcal{D}_j=\left\{\mathbf{x}^{(j)}, \mathbf{y}^{(j)}\right\}_{j}.
\end{equation}
{where $\mathbf{x}^{(j)}$ and $\mathbf{y}^{(j)}$ represent the vector form of $x(t)$ and $y(t)$ for the $j$-th class of channels, respectively. These DNN inputs are mapped to the desired output that is characterized by a specific number of taps.}

{Our problem is focused on finding a maximally sparse vector $\mathbf{c}^* \in \mathbb{R}^{m}$ such that the observed vector $\mathbf{r} \in \mathbb{R}^{n}$ can be represented by the fewest number of features in a feasible region. {In general,} the aforementioned problem is {considered an} optimization problem given by \cite{Xin2016} as}

\begin{equation}\begin{array}{cl}
\underset{\mathbf{c}}{\operatorname{min}} & \|\mathbf{c}\|_{0} \\
\text {s.t.} & \mathbf{r}=\mathbf{W} \mathbf{c}
\end{array}\end{equation}
{where $||\mathbf{c}||_0$ denotes $l^{0}$ norm, which counts the non-zero elements in $\mathbf{c}$,  reflects the number of channel taps. $\mathbf{W} \in \mathbb{R}^{n \times m}$ is {a known, overcomplete dictionary of feature vectors}. These vectors provide an indirect measurement for CIR.}

{The equivalent problem of (5) can be written as \cite{Xin2016}}
\begin{equation}\begin{array}{ll}
\underset{c}{\operatorname{min}} & \frac{1}{2}\|\mathbf{r}-\mathbf{W} \mathbf{c}\|_{2}^{2} \\
\text {s.t.} & \|\mathbf{c}\|_{0} \leqslant s
\end{array}\end{equation}
{where $s$ represents a pre-determined integer to control the sparsity of $\mathbf{c}$.}

If there is a strong coherence between columns of $\mathbf{W}$, then the sparse recovery estimation shown in (6) could be {significantly} poor.
{One possible solution to the observed high correlations in the dictionary is to exploit data for training a transformation of a dictionary, which can enhance its restricted isometry property constant.} 
{It has been proven that the existing DNNs provide such a solution \cite{Xin2016}.}


{Inspired by the featured characteristics offered by the DNN in \cite{Xin2016}, we employ a similar DNN, and its detailed structure can be found in Fig. \ref{networkfig}.} 
{To improve the performance of the used network, we incorporate some modifications {by introducing} two Dropout layers to the DNN proposed in \cite{Xin2016}.} {These Dropout layers offer regularization to avoid overfitting.}

{The adopted DNN includes fully-connected layers interleaved with non-linear transformation layers.} {In particular, the feed-forward design {comprises} fully-connected layers with 300 neurons and batch normalization.}
{The batch normalization is incorporated in the designed DNN to provide a reasonable initialization.
Simple Rectilinear units (ReLU) tend to promote sparse features by deactivating weights with negative values.} {The used non-linear unit can be represented as}
\begin{equation}
f(u)=u \operatorname{sign}(\max \{|u|-\alpha, 0\}),
\end{equation} 
{where sign(.) represents the sign function and $\alpha$ denotes the shrinkage parameter.}

Also, a final softmax layer is included, which yields a vector $\mathbf{p}$, where $p_j\in [0,1]$ {provides} an {assessment} of the likelihood that the {subject channel} belongs to the $j$-th class of channels. {For the loss function, we used the cross-entropy loss function, which is defined as} 

\begin{equation}
CE=-\sum_{j=1}^{C} g_{j} \log \left(p_{j}\right),
\end{equation}
where $C$ and $g_{j}$ represent the number of classes and ground truth for the $j$-th class, respectively, for a given observation.

\section{Methodology}
In this section, the datasets generation and simulation setup are presented.

\subsection{Training Data Generation} 
Here, {we import} the measurement-based channel datasets from the NYUSIM simulator \cite{NYUSIM2016}. This simulator is an open-source 5G and 6G channel model software. 
The existing channel simulators such as Quasi Deterministic Radio channel Generator (QuaDRiGa) \cite{quad2014}, Simulation of Indoor Radio Channel Impulse Response Models (SIRCIM) \cite{SIR1991}, etc. {have not been} developed based on extensive propagation measurements at centimeter-wave to millimeter-wave (mmWave) bands in diverse scenarios for 5G wireless systems. 
However, NYUSIM has been built based on extensive field measurements at mmWave bands in different outdoor environments, including urban microcell, urban macrocell, and rural macrocell environments.

Another {feature} found in NYUSIM is that it can recreate wideband power delay profiles/CIRs and channel statistics for {a} broad set of frequencies, beamwidths, bandwidths, wireless channel scenarios, etc. \cite{NYUSIM2016}. 
Moreover, the NYUSIM simulator generates a realistic three-dimensional statistical spatial CIRs {as represented in (2).} {These generated CIRs include the} clustering approach and physically-based {pathloss} model, which can be applied for the frequency range 0.5-100 GHz. 


\subsection{Simulation setup}
The parameters {of the conducted simulations} are {chosen to be} similar to that of the default simulation setup in \cite{NYUSIM2016} for {the} spatial channel model using NYUSIM. {The selected parameters are suitable in the 3rd Generation Partnership Project (3GPP) and other standard bodies as well as industrial/academic simulations.} 
The transmitted signals with a specific size of $1000\times 1$ are sent through different wideband frequency-selective channels with the generated CIR of {a} length corresponding to the number of multipath components. Then, the {complex-valued received signals} with their corresponding number of channel taps are used as {the} training dataset. 
{The inputs of the DNN are set as transmitted and received signals' samples. On the other hand, the output corresponds to the number of wireless channel taps.}
The received signal size is 1499, {where} real values are concatenated with its imaginary values. The dataset includes several {channel} realizations, limited to 222 different {channel} classes in terms of {the} number of channel taps. The DNN is implemented and trained using {the} Keras framework on Google Colaboratory service graphical processing unit (GPU).

{The specifications of the adopted DNN are defined as follows:} The chosen number of neurons is sufficiently large for {an adequate generalization performance}. However, this number should not be very large {not to} increase the computational complexity or cause overfitting. The training process is performed until a stop criterion is satisfied. 

The hyper-parameters selection is critical to the performance of DNN and they are determined empirically. 
The batch size is $128$ and the learning rate follows a scheduler. {The initial value of learning rate} is $0.001$ and decreases by a factor of 0.8 when the training accuracy forms a plateau for 18 epochs. Our DNN is trained with Adam optimizer \cite{Adam2014} due to its computational efficiency, fast, and smooth convergence.
The generated dataset size is $30$K observations which are divided into $70\%$ training data, $15\%$ validation data, and $15\%$ test data.

{The developed DNN is also compared with the existing SWISS algorithm} \cite{8377366}. 
{Table \ref{SwissPara} shows the simulation parameters of the SWISS algorithm}. It is proven in \cite{9107261} that choosing a proper threshold ($\eta$) becomes a more difficult problem as the number of paths increases, {which limits the achievable accuracy performance of the SWISS algorithm.} 

\begin{table}[!t]
\begin{center}
\caption{The considered parameters for the SWISS algorithm}
\resizebox{0.5\textwidth}{!}{
\begin{tabular}{|c|c|}
\hline
Threshold ($\eta$) & 0.995 \\ \hline
Number of pilots & 128 \\ \hline
Pilot energy & 1 \\ \hline
Modulation type & BPSK \\ \hline
Number of OFDM subcarriers & 512 \\ \hline
Number of OFDM symbols & 2 \\ \hline
Root-finding method & Newton's method \cite{kollerstrom_1992} \\ \hline
Initial guess in Newton's method & 1 \\ \hline
Number of iterations in Newton's method & 10 \\ \hline
Error tolerance in Newton's method & 0.001 \\ \hline
\end{tabular}
}
\label{SwissPara}
\end{center}
\end{table}

\section{Simulation Results}
For multi-label classification problem, the accuracy is {taken as one of the key performance metrics in the literature \cite{Xin2016}}. 
Fig. 2 {shows the convergence performance} after $1500$ training epochs. 
{The training accuracy is around $70\%$, about 8 points higher than the validation accuracy.} When the number of epochs increases, 
{both training and validation losses notably decrease.}
The achieved test accuracy is almost $62\%$. 
{These} preliminary results prove that our DNN converges to provide {significantly higher} accuracy {than the existing SWISS algorithm} under practical conditions where the signals are impaired by {the} frequency-selective fading {channels}.
{More specifically,} {the SWISS algorithm is a less efficient technique since pilot symbols are needed for accurate path-number identification.}
{The accuracy obtained from NYUSIM datasets using the SWISS algorithm is low (40.54\%), which is much less than the value obtained by the proposed DNN-based approach.}

\begin{figure}[!t]

             \begin{center}
              \subfloat[Change in training/validation accuracy during training.]{\label{convPerf:1}\includegraphics[width=45mm]{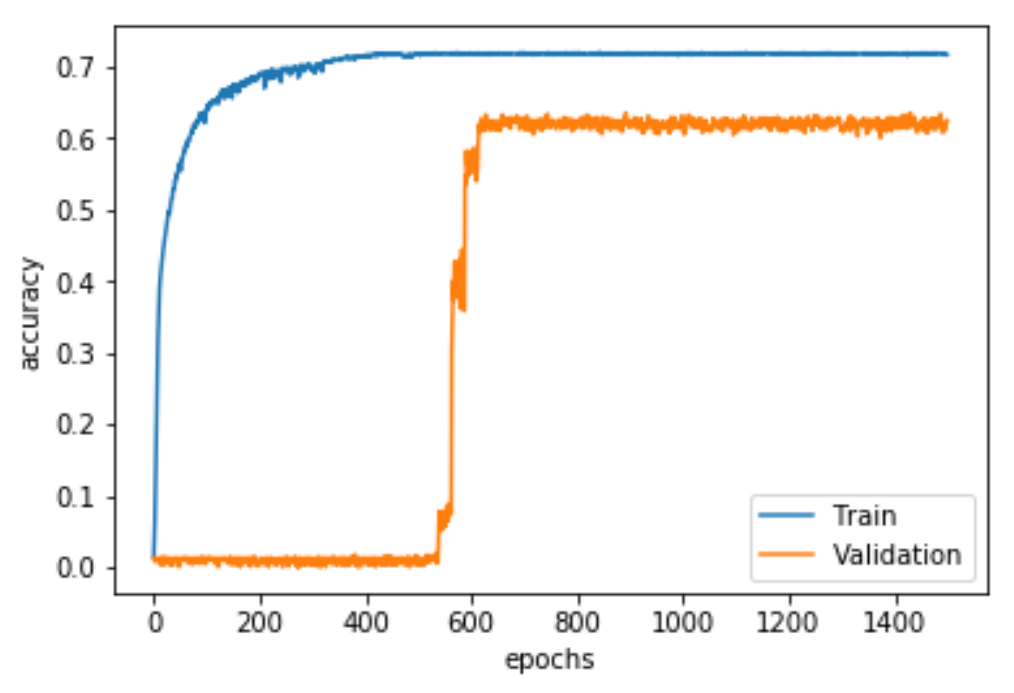}}
               \subfloat[Change in training/validation loss during training.]{\label{convPerf:2}\includegraphics[width=45mm]{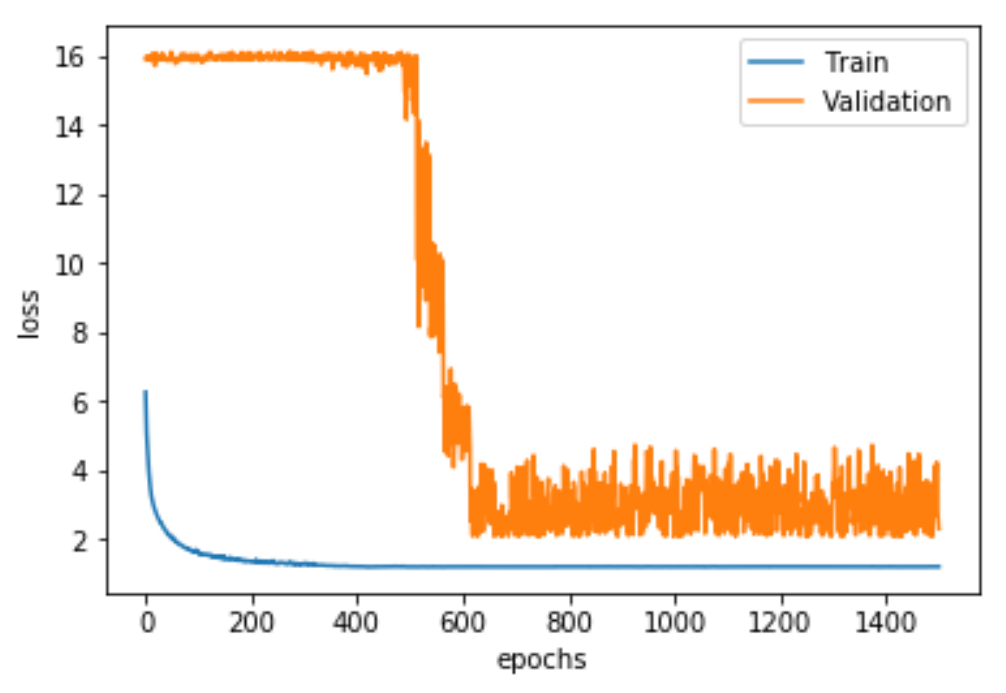}}
                \\
                \end{center}
                 \centering
           \label{convPerf}\caption{The convergence performance of the adopted DNN.}
\end{figure}

\begin{figure}[!t]

    \centering
    \includegraphics[height=1.7in]{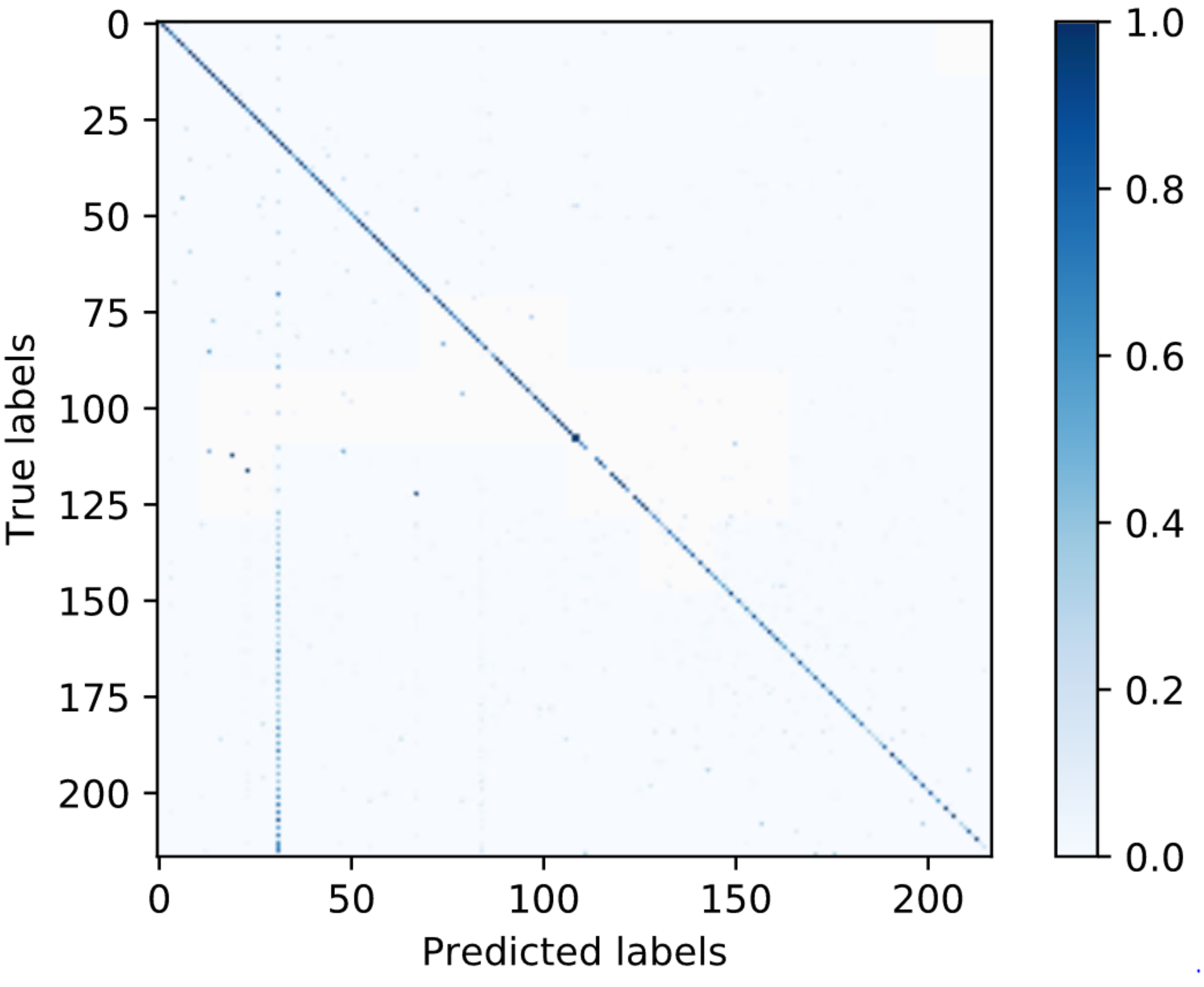}
    \caption{{Normalized confusion matrix for test set in the proposed DNN-based method.}}
    \label{conMatFig}
\end{figure}

{However, the accuracy metric can not represent the tolerance in results.}
The predicted number of taps that falls within a tolerance margin from the ground truth can be seen using {a} confusion matrix. The obtained confusion matrix for {the} test set is shown in Fig. \ref{conMatFig}.
However, the bias effect at the bottom left in Fig. \ref{conMatFig} can be explained by the nature of the channels generated from NYUSIM. These channels can have an unbalanced share of generation, causing the repetition of the number of taps in an unfair way. Thus, the bias leads to a misprediction that ultimately lowers the accuracy.
Despite {this unwanted effect}, the overall performance of the network tends to provide a solution {for the identification of} the number of channel taps.

{Furthermore, the computational complexity of the existing SWISS algorithm is very high.}
{Particularly, the SWISS technique requires implementing one of the root-finding methods such as the Newton method to get the optimal dual variable for each class, thereby increasing the computational overhead. Also, these root-finding methods need to have initial guesses, error tolerance, and several iterations to converge to the optimal dual variable.}
{Furthermore, the SWISS algorithm requires several iterations to reach the desired accuracy for each class.}
{On the other hand, the designed DNN saves computational complexity because it executes a constant number of matrix multiplications, independent of the channel's class.}

\section{Conclusion}
In this {paper}, {an end-to-end communication system is trained through a newly developed DNN to identify the number of channel taps.}
{We achieved an improved accuracy and loss performance over the existing SWISS algorithm} in identifying the number of channel taps. {Our DNN system is fed} by {a} measurement-based training dataset. 
The proposed method can be applied for different channel types without the need for a prior analysis. 
{Also, the identified number of taps could help track the time variations of the wireless channel caused by the Doppler effect.}
{The developed} DNN has the potential to identify not only {the} number of multipath {components but} also their corresponding locations. 
As another future direction, the impacts of the noise on the proposed solution {can} be analyzed.



\end{document}